\documentclass{aa}
\usepackage{graphicx}
\usepackage{txfonts}
\newcommand{\RSTAR}{\mbox{$R_{\star}$}}
\newcommand{\TEFF}{\mbox{$T_{\rm eff}$}}

\newcommand{\RSOL}{\mbox{$R_{\sun}$}}
\newcommand{\MSOL}{\mbox{$M_{\sun}$}}
\newcommand{\LSOL}{\mbox{$L_{\sun}$}}
\newcommand{\micron}{\mbox{$\mu$m}}
\newcommand{\KMS}{\mbox{km s$^{-1}$}}
\newcommand{\RIN}{\mbox{$r_{\rm in}$}}
\newcommand{\ROUT}{\mbox{$r_{\rm out}$}}
\newcommand{\HOH}{\mbox{H$_2$O}}
\newcommand{\TAUV}{\mbox{$\tau_{V}$}}
\newcommand{\MCSIM}{\mbox{\sf mcsim\_mpi}}
\newcommand{\AMIN}{\mbox{$a_{\rm min}$}}
\newcommand{\AMAX}{\mbox{$a_{\rm max}$}}
\newcommand{\MBOL}{\mbox{$M_{\rm bol}$}}
\begin{document}
\title{
Spatially resolved dusty torus toward the red supergiant WOH G64 in the 
Large Magellanic Cloud
\thanks{
Based on observations made with the Very Large Telescope Interferometer of 
the European Southern Observatory. Program ID: 076.D-0253, 080.D-0222}
\fnmsep
\thanks{
This work is based [in part] on observations made with the Spitzer Space 
Telescope, which is operated by the Jet Propulsion Laboratory, California 
Institute of Technology under a contract with NASA.
}
}
\subtitle{
}

\author{K.~Ohnaka\inst{1} 
\and
T.~Driebe\inst{1} 
\and
K.-H.~Hofmann\inst{1} 
\and
G.~Weigelt\inst{1} 
\and
M.~Wittkowski\inst{2}
}

\offprints{K.~Ohnaka}

\institute{
Max-Planck-Institut f\"{u}r Radioastronomie, 
Auf dem H\"{u}gel 69, 53121 Bonn, Germany\\
\email{kohnaka@mpifr-bonn.mpg.de}
\and
European Southern Observatory, Karl-Schwarzschild-Str.~2, 
85748 Garching, Germany
}

\date{Received / Accepted }

\abstract
{}
{We present $N$-band spectro-interferometric observations of the red 
supergiant WOH G64 in the Large Magellanic Cloud (LMC) using MIDI at 
the Very Large Telescope Interferometer (VLTI).  
While the very high luminosity ($\sim \! 5 \times 10^5$~\LSOL) previously 
estimated for WOH G64 suggests that it is a very massive star with an 
initial mass of $\sim$40~\MSOL, its low effective temperature ($\sim$3200~K) 
is in serious disagreement with the current stellar evolution theory.  
}
{
WOH G64 was observed with VLTI/MIDI using the UT2-UT3 and UT3-UT4 
baseline configurations. 
}
{
The dust envelope around WOH G64 has been spatially resolved with a 
baseline of $\sim$60~m---the first MIDI observations to resolve an 
individual stellar source in an extragalactic system.  
The observed $N$-band visibilities show a slight decrease from 8 to 
$\sim$10~\micron\ and a gradual increase longward of $\sim$10~\micron, 
reflecting the 10~\micron\ silicate feature in self-absorption.  
This translates into a steep increase of the uniform-disk diameter from 8 to 
10~\micron\ (from 18 to 26~mas) and a roughly constant diameter above 
10~\micron.  
The visibilities measured at four position angles differing by $\sim$60\degr\ 
but at approximately the same baseline length ($\sim$60~m) do not show 
a noticeable difference, suggesting that the object appears nearly 
centrosymmetric.  
The observed $N$-band visibilities and spectral energy distribution 
can be reproduced by an optically and geometrically thick silicate torus 
model viewed close to pole-on. 
The luminosity of the central star is derived to be 
$\sim \! 2.8 \times 10^5$~\LSOL, which is by a factor of 2 lower than the 
previous estimates based on spherical models.  
We also identify the \HOH\ absorption features at 2.7 and 6~\micron\ in 
the spectra obtained with the Infrared Space Observatory and 
the Spitzer Space Telescope.  
The 2.7~\micron\ feature originates in the photosphere and/or the extended 
molecular layers, 
while the 6~\micron\ feature is likely to be of 
circumstellar origin. 
}
{
The lower luminosity newly derived from our MIDI observations and 
two-dimensional modeling 
brings the location of WOH G64 on the H-R diagram 
in much better agreement with theoretical evolutionary tracks 
for a 25~\MSOL\ star.  However, the effective temperature is still 
somewhat too cool compared to the theory.  
The low effective temperature of WOH G64 places it very close to or even 
beyond the Hayashi limit, which implies that this object may be experiencing 
unstable, violent mass loss.  
}


\titlerunning{Spatially resolved dusty torus toward the LMC red supergiant WOH G64}
\authorrunning{Ohnaka et al.}
\maketitle

\section{Introduction}
\label{sect_intro}

The evolution of massive ($M \ga 8 \, M_{\odot}$) red supergiants (RSGs) 
is not well understood, because of several physical processes 
difficult to theoretically formulate, such as mass loss, 
convective mixing, and rotation (e.g., Massey \cite{massey03}).  
Comparison of theoretical evolutionary tracks for RSGs with 
observational data is crucial for testing the present theory of 
the evolution of massive stars. 
While it is usually difficult to reliably derive the luminosities of 
Galactic RSGs, their counterparts 
in the Large and Small Magellanic Clouds (LMC and SMC, 
respectively) provide us with an excellent opportunity to compare 
theory and observation on the H-R diagram, thanks to their known 
distances ($\sim$50 and $\sim$60~kpc for the 
LMC and SMC, respectively).  
Another advantage of studying RSGs in the LMC and SMC is that we can 
probe possible metallicity effects on the mass loss, because the 
metallicities of the LMC and SMC are systematically lower than the 
solar value ($\sim \! 0.5 \, Z_{\sun}$ and $0.2 \, Z_{\sun}$
for the LMC and SMC, respectively).  
Although radiation pressure on dust grains is often considered to be the 
driving mechanism of mass loss in cool evolved stars, 
it is little understood where and how dust forms in RSGs and 
how mass outflows are initiated.  
Since dust formation is expected to depend on metallicity, studies on 
the circumstellar environment of RSGs in the LMC and SMC are useful 
for clarifying the metallicity effect on the dust formation 
and obtaining insights into the driving mechanism of mass outflows.

\begin{figure}[!t]
\resizebox{\hsize}{!}{\rotatebox{0}{\includegraphics{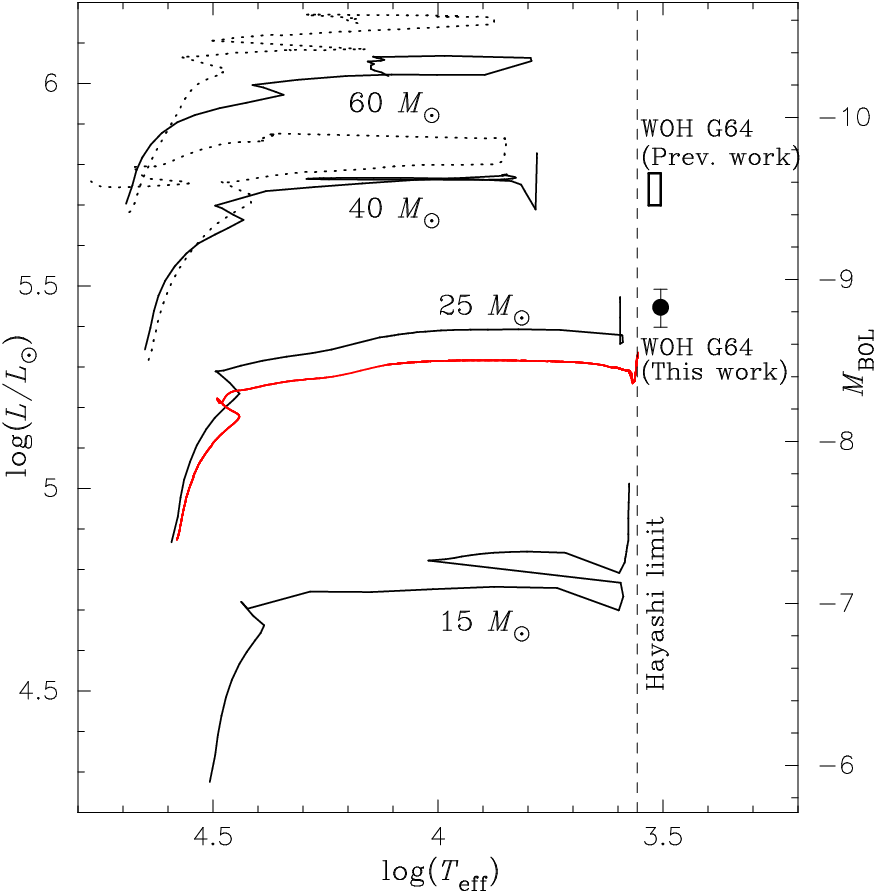}}}
\caption{H-R diagram with theoretical evolutionary tracks and the 
observationally derived locations of WOH G64.  
The black solid lines represent the evolutionary tracks for stars 
with 15, 25, 40, and 60~\MSOL\ with $Z = 0.008$ from 
Schaerer et al. (\cite{schaerer93}).  
The evolutionary tracks with an initial rotation velocity of 300~\KMS\ 
for $Z = 0.008$ from Meynet \& Maeder (\cite{meynet05}) 
are plotted with the dotted lines, if available.  
The red solid line represents the newly computed track for a 25~\MSOL\ 
with $Z = 0.01$ with the code of one of the authors (T.~Driebe).  
The Hayashi limit presented in Levesque et al. (\cite{levesque07}) is 
shown with the dashed line.  
The location of WOH G64 based on \TEFF\ = 3200--3400~K and 
(5--6)$\times 10^5$~\LSOL\ derived in the previous studies is shown
with the box.  
The filled circle represents the location of WOH G64 based on the luminosity 
of $2.8 \times 10^5$~\LSOL\ derived in the present work (see 
Sect.~\ref{sect_discuss}).  
}
\label{hr_diagram}
\end{figure}

WOH G64 (IRAS04553-6825) is a highly luminous cool star in the LMC first 
identified by Westerlund et al. (\cite{westerlund81}).  
Subsequent infrared observations 
by Elias et al. (\cite{elias86}) and Roche et al. (\cite{roche93}) 
revealed that it exhibits a huge infrared excess and the 10~\micron\ 
silicate feature in self-absorption, which means that the object is 
surrounded by an optically thick dust envelope, experiencing heavy mass 
loss.  
WOH G64 is also known to show OH, SiO, and \HOH\ masers (Wood et al. 
\cite{wood86}, \cite{wood92}; van Loon et al. \cite{vanloon96}, 
\cite{vanloon98}, \cite{vanloon01}; Marshall et al. \cite{marshall04}). 
The spectroscopic studies on the TiO absorption bands suggest  
spectral types of M5--M7 (Elias et al. \cite{elias86}; Van Loon et al. 
\cite{vanloon05}), which corresponds to effective temperatures of 3200--3400~K 
(see Sect.~\ref{sect_modeling}).  
Elias et al. (\cite{elias86}) and van Loon et al. (\cite{vanloon05}) derived 
bolometric magnitudes of $-9.7$ ($6 \times 10^5$~\LSOL) and $-9.4$ 
($5 \times 10^5$~\LSOL), respectively, assuming spherical dust shells.  
Figure~\ref{hr_diagram} shows an H-R diagram with the evolutionary models 
for $Z = 0.008$ of Schaerer et al. (\cite{schaerer93}).  For comparison, 
we also plot the evolutionary track for a 25~\MSOL\ star with $Z = 0.01$ 
newly computed with another evolutionary code (author: T.~Driebe), which is 
based on the program of Bl\"ocker (\cite{bloecker95}).  
The convective overshoot is included as described in 
Herwig et al. (\cite{herwig97}) and Herwig (\cite{herwig00}) with an 
overshoot parameter of $f = 0.004$, and the mass loss is incorporated 
by using the formula of de Jager et al. (\cite{dejager88}) from the zero-age 
main sequence.  
Figure~\ref{hr_diagram} suggests that a star with an initial mass of 
$\sim$40~\MSOL\ can reach the high luminosities observationally 
derived for WOH G64.  
However, the evolutionary models predict that a 40~\MSOL\ star evolves 
only down to $\sim$6300~K and does not reach the low effective temperatures of 
3200--3400~K observed for WOH G64.  
Therefore, the location of WOH G64 on the H-R diagram is in serious 
disagreement with theory.

It is worth noting here that Roche et al. (\cite{roche93}) and van Loon 
et al. (\cite{vanloon99}) suggest a disk-like or clumpy structure for the 
dust envelope of WOH G64.  Van Loon et al. (\cite{vanloon99}) 
attempted to reproduce the observed spectral energy distribution (SED) 
with a spherical dust shell model, but could not find a reasonable fit.  
The reason is that the huge mid- and far-infrared excess indicates a 
large optical depth for the dust shell, which would lead to very low 
flux in the near-infrared and in the visible.  However, the observed 
SED of WOH G64 shows high fluxes at these wavelengths. 
The possible presence of a disk-like or clumpy structure makes the above 
luminosity estimates assuming spherical dust shells uncertain.  
If there is indeed such a deviation from spherical symmetry, 
the luminosity of the central star can only be 
estimated from multi-dimensional radiative transfer modeling using 
as many observational constraints as possible.  
Therefore, observational studies on the disk-like/clumpy structure in 
WOH G64 is important both for deriving the luminosity of the central star 
and for a better understanding of the mass loss in RSGs in the LMC.

High spatial resolution observations in the mid-infrared are a powerful tool 
to obtain a clearer picture of the circumstellar environment of RSGs.  
The MID-infrared Interferometric instrument (MIDI) at the Very Large Telescope 
Interferometer (VLTI) is ideal for this goal with its 
``spectro-interferometric'' capability and has already proven 
its potential to study the circumstellar environment of cool (and also hot) 
evolved stars.  
However, such studies on stellar sources with MIDI are limited to Galactic 
objects up to now.  
Despite the large distance to the LMC (50~kpc is adopted in the present 
work), the high luminosity of WOH G64 and its huge infrared excess make it 
observable with MIDI ($F_{12\mu{\rm m}} = 7$~Jy).  
In this paper, we present MIDI observations of WOH G64 and radiative 
transfer modeling to derive the physical properties of its dust envelope 
as well as the luminosity of the star.

\section{Observations}
\label{sect_obs}

\begin{table*}[t]
\begin{center}
\caption {Summary of the MIDI observations of WOH G64: night, time of 
observation (Coordinated Universal Time=UTC), telescope configuration (Tel.), 
projected baseline length $B_{\rm p}$, position angle of the projected 
baseline on the sky (P.A.), and seeing.  
The data sets \#1 and \#3 were used only for extracting the $N$-band spectra.  
}
\vspace*{-2mm}

\begin{tabular}{r c c c r r l l}\hline
\# & Night  & $t_{\rm obs}$ & Tel. & $B_{\rm p}$ & P.A.   & Seeing & Remarks\\ 
   &        & (UTC)         &      & (m)         & (\degr)&        &        \\
\hline
1  & 2005 Sep. 18 & 08:18:03 & UT3-UT4 & 57.9 & 85.7  &  1\farcs5 & 
Spectrum only \\
2  & 2005 Nov. 12 & 08:22:17 & UT3-UT4 & 62.4 & 136.0 & 0\farcs5--0\farcs7 & \\
3  & 2005 Nov. 13 & 08:23:15 & UT2-UT3 & 33.9 & 77.7  & 0\farcs6  &
Spectrum only\\
4  & 2007 Oct. 22 & 05:30:49 & UT3-UT4 & 56.5 & 77.3  & 1\farcs0--1\farcs5 & \\
5  & 2007 Oct. 22 & 06:36:17 & UT3-UT4 & 59.0 & 92.7  & 1\farcs0--1\farcs5 & \\
6  & 2007 Oct. 22 & 07:41:05 & UT3-UT4 & 60.8 & 107.5 & 1\farcs0--1\farcs5 & \\
\hline
\label{table_obs}
\vspace*{-7mm}

\end{tabular}
\end{center}
\end{table*}

\begin{figure}[!hbt]
\resizebox{\hsize}{!}{\rotatebox{0}{\includegraphics{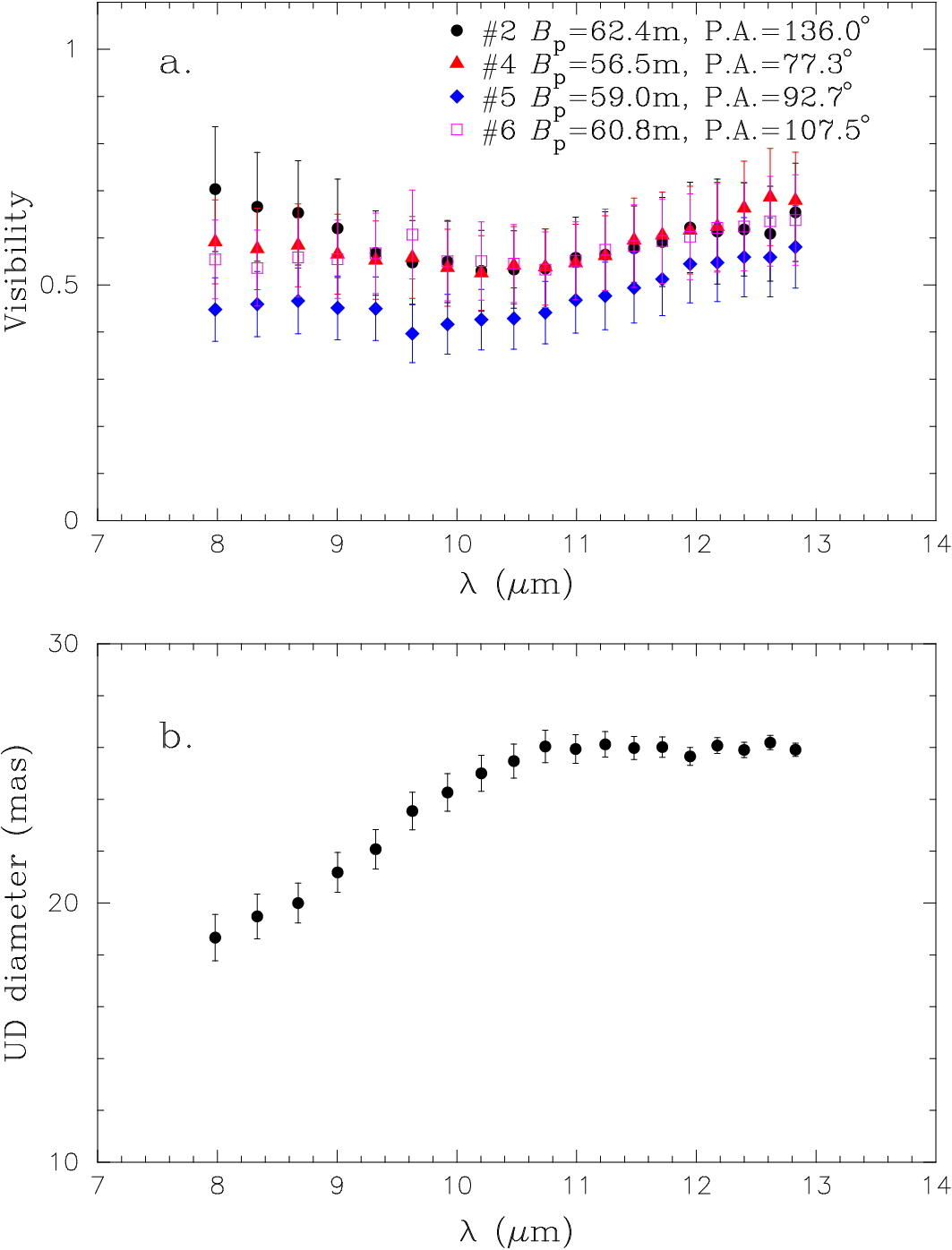}}}
\caption{{\bf a:} $N$-band visibilities of WOH G64 derived from the 
data sets \#2, \#4, \#5, and \#6.  
The errors are dominated by the systematic errors as described in 
Sect.~\ref{sect_obs}.  
{\bf b:} Uniform-disk (UD) diameter obtained by fitting the observed 
visibilities. 
}
\label{obs_vis}
\end{figure}

\subsection{$N$-band visibilities}

WOH G64 was observed with MIDI in the HIGH\_SENS mode 
in September and November 2005 as well as in October 2007 
with 
two baseline configurations using the 8.2~m Unit Telescopes (UT2-UT3 and 
UT3-UT4), as summarized in Table~\ref{table_obs} 
(Program ID: 076-D0253 and 080-D0222, P.I.: K.~Ohnaka).  
A prism with a spectral resolution of $\lambda/\Delta \lambda \simeq 30$ 
at 10~\mbox{$\mu$m}\ was used to obtain spectrally dispersed fringes 
between 8 and 13~\mbox{$\mu$m}, and 
each data set contains typically 200 scans.  
A detailed description of the observing procedure is given in 
Przygodda et al. (\cite{przygodda03}), Leinert et al.\ (\cite{leinert04}), 
and Chesneau et al. (\cite{chesneau05}).  
Since WOH G64 itself is too faint in the $V$ band ($V \ga 17$) 
for the Multi-Application Curvature sensing Adaptive Optics 
(MACAO) system, we performed ``off-target Coud\'e guiding'' using the nearby 
star 0216-0046676 (USNO-B1.0 catalog) with $V$ = 16, which is 10\farcs6 away 
from WOH G64.  
In addition to MACAO, the InfraRed Image Sensor (IRIS) operating in the 
$K$ band was used to keep the alignment of two beams as correct as possible 
during interferometric observations.  

For data reduction, 
we use the MIA+EWS package ver.1.5\footnote{Available at 
http://www.strw.leidenuniv.nl/\textasciitilde nevec/MIDI} 
(Leinert et al. \cite{leinert04}; Jaffe \cite{jaffe04}).  
Since the $V$ magnitude of the guide star 
is close to the limit of the MACAO system on UTs ($V$ = 17), our MIDI 
observations can be severely affected by atmospheric conditions such as 
seeing, wind speed, and coherence time.  
In order to assess the data quality, we check the acquisition images 
of each data set, which are taken before the interferometric observations, 
as well as the histogram of the fringe power of scans 
at each spectral channel.  This fringe histogram appears very broad 
and/or asymmetric when the image quality is poor or the overlap 
of two beams is not perfect, and therefore, it can be used for diagnosing 
the data quality.  
We find out that the data sets \#1 and \#3 are affected by poor image 
quality and/or imperfect beam overlap, and therefore, we use these data 
sets only for extracting the $N$-band spectra.  
On the other hand, the other four data sets do not show a hint of such 
problems, and we derive visibilities and spectra from these data sets.  

We derive the interferometer transfer function at each spectral channel 
between 8 and 13~\mbox{$\mu$m}\ by observing calibrators listed in 
Table~\ref{table_calib}. 
We use the mean of the transfer function values derived from the 
calibrators observed on the same night as WOH G64, and 
the errors of the calibrated visibilities are first estimated from the 
standard deviation of the transfer function values at each wavelength.  
While the errors estimated in this manner are quite small (1$\sigma$ 
$\approx 5$\%), 
there may still be systematic errors discussed above, even if the 
data sets seem not to be severely affected by the overlap problem and/or 
poor image quality.  In order to account for such systematic 
errors, we assume a minimum relative error of 15\% as adopted by 
Preibisch et al. (\cite{preibisch06}).

Figure~\ref{obs_vis}a shows the $N$-band visibilities of WOH G64 derived 
from the data sets \#2, \#4, \#5, and \#6 using the EWS package.  
Since the results obtained with the MIA and EWS packages show good agreement, 
we only present the result derived with the EWS package in the discussion 
below.  
The object is spatially resolved by our MIDI observations.  The observed 
$N$-band visibilities are approximately constant or slightly decrease 
from 8 to $\sim$10~\micron, while they increase gradually longward of 
10~\micron. 

Qualitatively, the observed wavelength dependence of the visibilities can be 
explained by the silicate absorption feature at 10~\micron\ 
(see Fig.~\ref{obs_spec}).   In an optically thick dust envelope, photons in 
the silicate feature originate farther out from the star compared to the 
continuum, leading to a decrease in visibility inside the silicate 
feature.  
The four visibilities were obtained at approximately the same 
baseline length (57--62~m) but at position angles differing by 59\degr, but 
there is no remarkable difference among the visibilities within the 
errors.  
Therefore, the object shows no significant deviation from centrosymmetry 
over this range of position angle.  
The uniform-disk diameter derived from the visibilities, which is plotted in 
Fig.~\ref{obs_vis}b, shows a steep increase from 8 to 10~\micron\ (from 18 to 
26~mas) with a flat portion longward of 10~\micron.  
However, we stress that the uniform-disk fitting is merely for obtaining 
some characteristic angular size of the object.

\subsection{$N$-band spectra}

\begin{figure}[!hbt]
\resizebox{\hsize}{!}{\rotatebox{-90}{\includegraphics{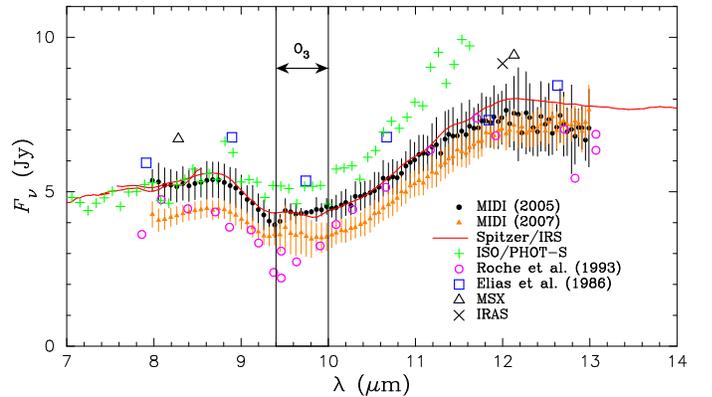}}}
\caption{$N$-band spectra of WOH G64 observed with MIDI, plotted together 
with the Spitzer/IRS and ISO/PHOT-S spectra as well as (spectro)photometric 
data available in the literature.  The region severely affected by the 
ozone band (9.6--10~\micron) is marked by the vertical lines. 
}
\label{obs_spec}
\end{figure}

We extract the absolutely calibrated $N$-band spectra of WOH G64 
from our MIDI data with the procedure described in Ohnaka et al. 
(\cite{ohnaka07}).  The calibrators observed at airmasses similar to 
WOH G64 are used for the spectrophotometric calibration and marked with 
$\dagger$ in Table~\ref{table_calib}.  
The averaged spectra are calculated separately from the data taken 
2005 and 2007 with three (WOH G64, calibrator) pairs each, and 
the errors are estimated from the standard deviation of the spectra 
derived from these pairs.  
The extracted $N$-band spectra, which are plotted in Fig.~\ref{obs_spec}, 
are dominated by the self-absorption of amorphous silicate at 10~\micron.  
The MIDI spectrum obtained in 2007 shows flux levels slightly lower than 
that observed in 2005.  However, given the errors of the spectrophotometric 
calibration of MIDI data, the difference is marginal, and we cannot draw 
a definitive conclusion about the temporal variation of the $N$-band flux 
between 2005 and 2007.  
Also plotted are the mid-infrared spectra obtained with the Spitzer Space 
Telescope (Werner et al. \cite{werner04}) and the Infrared Space Observatory 
(ISO).  
The Spitzer spectrum was obtained on 2005 January 12 with the InfraRed 
Spectrometer\footnote{
The IRS was a collaborative venture between Cornell University and Ball 
Aerospace Corporation funded by NASA through the Jet Propulsion Laboratory 
and Ames Research Center.
} (IRS, Houck et al. \cite{houck04}) in the Short Low (SL) and 
Long Low (LL) modes covering from 5.2 to 38~\micron\ with spectral resolution 
of 64--128 (Program ID: P03426, P.I.: J. H. Kastner) 
and has recently been published by Buchanan et al. (\cite{buchanan06}).  
We downloaded the Basic Calibration Data (BCD) processed with the S15-3 
pipeline from the Spitzer data archive and extracted the spectrum using 
SMART v.6.2.4 (Higdon et al. \cite{higdon04}).  
The sky background was estimated by means of ``local sky'' 
using the pixels outside the extraction aperture, and the spectra extracted 
at two nodding positions are co-added.  The extracted spectrum 
is in very good agreement with that published in 
Buchanan et al. (\cite{buchanan06}).  
The ISO spectrum obtained on 1996 May 26 with PHOT-S (2.5--5~\micron\ and 
6--12~\micron, $\lambda/\Delta \lambda \approx 90$), which was published 
by Trams et al. (\cite{trams99}), was downloaded from the ISO data archive.  
We also plot (spectro)photometric data available in the literature 
(Elias et al. \cite{elias86}; Roche et al. \cite{roche93}; 
MSX, Egan et al. \cite{egan03}; IRAS Point Source Catalog).  

The MIDI spectrum obtained in 2005 
is in very good agreement with the Spitzer/IRS spectrum, 
which was taken relatively close in time to our MIDI data compared 
to the other data.  
We also note that the MIDI spectrum was obtained with a slit of 
0\farcs5$\times$2\arcsec, while the Spitzer/IRS uses much larger slits 
(3\farcs6 $\times$57\arcsec\ and 10\farcs5 $\times$168\arcsec\ in 
the SL and LL modes, respectively).  Therefore, the agreement 
between the MIDI and Spitzer/IRS spectra means the absence of extended 
emission in the $N$ band (whether physically associated to WOH G64 or 
background emission). 
On the other hand, 
there are some differences in the flux level as well 
as the depth of the 10~\micron\ feature between the MIDI spectra 
and the other data.  This can be due to either differences in the apertures 
used for these observations or an intrinsic temporal variation.  
Given the above good agreement between the MIDI and Spitzer data,  
the differences between the MIDI and the other data are more likely to be 
(long-term) temporal variations of the mid-infrared flux.  
In fact, Roche et al. (\cite{roche93}) report that WOH G64 shows noticeable 
temporal variations of the flux level as well as the depth of the 10~\micron\ 
absorption over $\sim$5~years, 
despite the small variability amplitude in the $K$ band 
($\Delta K \approx 0.3$, Wood et al. \cite{wood92}; 
Whitelock et al. \cite{whitelock03}).  
However, since the MIDI spectra taken in 2005 and 2007 do not show a 
significant temporal variation, we merge the data ($N$-band visibilities 
as well as spectra) taken at the two epochs for our radiative transfer 
modeling.  In addition to our MIDI spectra, we also include the Spitzer/IRS 
spectrum in our modeling, 
because it was obtained relatively close in time to 
our MIDI observations.

\begin{table}[h]
\begin{center}
\caption {
List of calibrators used in the present work, 
together with their spectral type, 12~\mbox{$\mu$m}\ flux ($F_{12}$), 
uniform-disk diameter ($d_{\rm{UD}}$) 
and the date as well as the time stamp ($t_{\rm obs}$) 
of the MIDI observations. 
The uniform-disk diameters were taken from the CalVin 
list available at ESO
(http://www.eso.org/observing/etc/). 
The stars used for spectrophotometric calibration are marked with $\dagger$. 
}
\begin{tabular}{l l r r c l}\hline
Calibrator & Sp.  & $F_{12}$ & $d_{\rm{UD}}$ & Date    & $t_{\rm obs}$ \\ 
           & Type   & (Jy)  & (mas)          &         & (UTC)         \\ \hline
\multicolumn{6}{c}{2005}\\
\hline
HD33042    & K5III  & 12.3  & $2.79\pm 0.15$ & Sep. 18 & 07:37:50${\dagger}$ \\ 
           &        &       &                & Nov. 12 & 08:46:31      \\
           &        &       &                & Nov. 13 & 08:47:36${\dagger}$ \\
HD37160    & K0IIIb & 9.4   & $2.08\pm 0.20$ & Sep. 18 & 08:56:35${\dagger}$ \\
HD36673    & F0Ib   & 8.1   & $1.51\pm 0.27$ & Nov. 12 & 05:33:39      \\
           &        &       &                & Nov. 12 & 07:10:25      \\
HD36079    & G5II   & 20.3  & $2.97\pm 0.16$ & Nov. 12 & 06:24:46      \\ \hline
\multicolumn{6}{c}{2007}\\
\hline
HD26967    &  K2III & 13.5  & $2.59\pm 0.15$ & Oct. 22 & 04:56:51  \\
HD55865    &  K0III & 12.4  & $2.49\pm 0.13$ & Oct. 22 & 06:08:55  \\
           &        &       &                & Oct. 22 & 07:05:45${\dagger}$  \\
\hline
\label{table_calib}
\end{tabular}
\end{center}
\end{table}

\begin{figure}
\resizebox{\hsize}{!}{\rotatebox{0}{\includegraphics{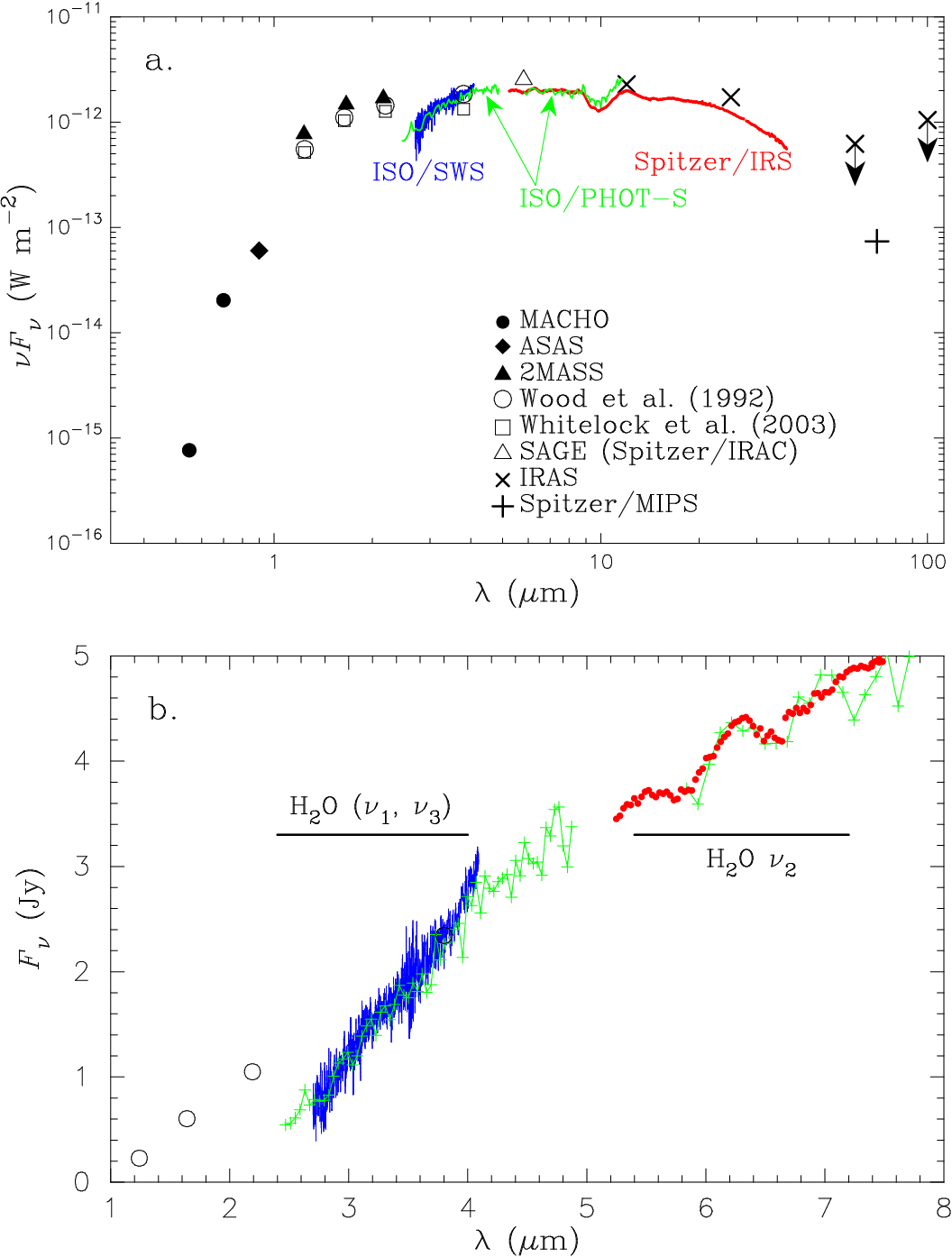}}}
\caption{{\bf a:} SED of WOH G64 from the optical to the far-infrared.  
The photometric data are taken from the literature and various catalogs 
as given in the panel.  The data from Wood et al. (\cite{wood92}) and 
Whitelock et al. (\cite{whitelock03}) are the time-averaged values.  
Only the 5.8~\micron\ flux measured with the InfraRed Array Camera 
(IRAC) of Spitzer is currently available in the SAGE database.  
The IRAS 60~\micron\ and 100~\micron\ fluxes represent upper limits.  
{\bf b:} The \HOH\ absorption features in the region between 1 and 8~\micron. 
Solid line (between 2.7 and 4~\micron): ISO/SWS.  Crosses: ISO/PHOT-S.  
Filled circles: Spitzer/IRS.  Open circles: Photometric data from 
Wood et al. (\cite{wood92}).  
}
\label{obs_sed}
\end{figure}

\subsection{Spectral energy distribution}
\label{subsect_sed}

Together with the $N$-band visibilities and mid-infrared MIDI and Spitzer/IRS 
spectra, we use photometric data from the optical to the far-infrared 
as observational constraints in our radiative transfer modeling.  
We collect photometric data on WOH G64 available in the literature 
(Elias et al. \cite{elias86}; Wood et al. \cite{wood92}; 
Whitelock et al. \cite{whitelock03}; 
2MASS, Cutri et al. \cite{cutri03}; 
MACHO, Alcock et al. \cite{alcock01}; 
ASAS, Pojmanski et al al. \cite{pojmanski02}, \cite{pojmanski03}, 
\cite{pojmanski04}, \cite{pojmanski05a},b; 
IRAS Point Source Catalog; 
SAGE, Meixner et al. \cite{meixner06}).  
The long-term monitoring of Wood et al. (\cite{wood92}) and 
Whitelock et al. (\cite{whitelock03}) shows that the variability of WOH G64 
is as small as $\sim$0.3 in the near-infrared.  
The infrared spectra obtained with the PHOT-S and the Short 
Wavelength Spectrometer (SWS) onboard the ISO (Vandenbussche et al. 
\cite{vandenbussche02}) are also included.  
Since the IRAS data only give upper limits on the 60 and 100~\micron\ fluxes, 
we derive the 70~\micron\ flux from the imaging data obtained with the 
Multiband Imaging Photometer for Spitzer (MIPS, Rieke et al. \cite{rieke04}) 
on 2005 November 8 (Program ID: P20203, P.I.: M. Meixner).  
The 70~\micron\ flux is extracted from the 
post-BCD data with an aperture 
radius of 35\arcsec, with the background emission estimated in an annulus 
between 40\arcsec\ and 60\arcsec.  The resulting 70~\micron\ flux is 1.7~Jy. 
The interstellar extinction toward WOH G64 is uncertain.  
Van Loon et al. (\cite{vanloon97}) argue that while the foreground reddening 
toward the LMC is $A_V \approx 0.18$ on average, there may be considerable 
(but patchy) extinction inside the LMC.  
However, since no systematic study on the 
reddening inside the LMC is available, we neglect the interstellar extinction 
toward WOH G64 in the present work.  
The whole SED is plotted in Fig.~\ref{obs_sed}a, which reveals the huge
infrared excess toward WOH G64.  

Figure~\ref{obs_sed}b shows an enlarged view of the region between 1 and 
8~\micron.  The ISO/SWS and ISO/PHOT-S spectra reveal remarkable absorption 
features between 2.4 and 4~\micron, which can be attributed to the \HOH\ 
$\nu_1$ and $\nu_3$ fundamental bands.  
As presented in Sect.~\ref{sect_modeling}, 
the flux contribution of the attenuated star is the more dominant 
compared to the 
dust thermal emission at wavelengths below $\sim$3~\micron.  Therefore, 
the \HOH\ absorption feature centered at 2.7~\micron\ is likely to 
originate in the photosphere and/or in the 
so-called ``warm molecular layers'' (or MOLsphere) close to the star, 
whose presence is confirmed in Galactic RSGs (e.g., Tsuji \cite{tsuji00}, 
\cite{tsuji06}; Ohnaka \cite{ohnaka04}; 
Verhoelst et al. \cite{verhoelst06}; Perrin et al. \cite{perrin07}).  
On the other hand, the Spitzer/IRS spectrum (and also the ISO/PHOT-S 
spectrum, although a little noisy) reveals the 6~\micron\ \HOH\ feature due to 
the $\nu_2$ fundamentals, which is only marginally discernible in Fig.~1a 
of Buchanan et al. (\cite{buchanan06}) because of the large ordinate scale 
of their plot.  
Since the flux is dominated by the dust thermal emission at these wavelengths, 
the 6~\micron\ absorption feature is mostly caused by the \HOH\ gas 
present in the circumstellar envelope, and the \HOH\ absorption is seen 
with the hot dust thermal emission as the background.  
Such circumstellar \HOH\ $\nu_2$ absorption features, albeit much weaker, 
are also identified in the Galactic red supergiant NML~Cyg (Justtanont et al. 
\cite{justtanont96}).

\section{Modeling of the dusty environment of WOH G64}
\label{sect_modeling}

Since van Loon et al. (\cite{vanloon99}) have already shown that the SED of 
WOH G64 cannot be explained by a spherical dust shell model, we attempt 
to explain the observed SED and $N$-band visibilities with the second 
simplest geometry: axisymmetric dust torus (or disk) model.  
We use our Monte Carlo code (\MCSIM) for two-dimensional modeling of 
the dusty environment of WOH G64.  
The outputs of \MCSIM\ are 
dust temperature and monochromatic mean intensity in each cell of the 
dust envelope.  Using these outputs, a model image at a given wavelength 
viewed from an arbitrary angle is computed with our ray-trace code.  
The details of our codes are described in Ohnaka et al. (\cite{ohnaka06}). 

The central star is assumed to be a blackbody of effective temperature 
\TEFF.  
The spectral type of WOH G64 is determined as M7.5 from optical 
spectroscopy by Elias (\cite{elias86}) and van Loon et al. (\cite{vanloon05}), 
although the latter authors note some ambiguity in the spectral classification 
of WOH G64 and suggest that an earlier spectral type of M5 is also possible.  
We convert these spectral types to effective temperatures using the 
temperature 
scale derived by Levesque et al. (\cite{levesque06}) based on RSGs 
in the LMC.  Since their effective temperature scale does not extend beyond 
M4.5, we linearly extrapolate the temperature scale between M0 and M4.5 down 
to M7.5, which results in \TEFF\ = 3230~K.   The earlier spectral type of M5 
corresponds to \TEFF\ = 3400~K.  We adopt \TEFF\ = 3200~K in our modeling, 
but the adoption of 3400~K turns out not to noticeably affect the result 
of the modeling presented below.  

\begin{figure}[!hbt]
\resizebox{8.5cm}{!}{\rotatebox{0}{\includegraphics{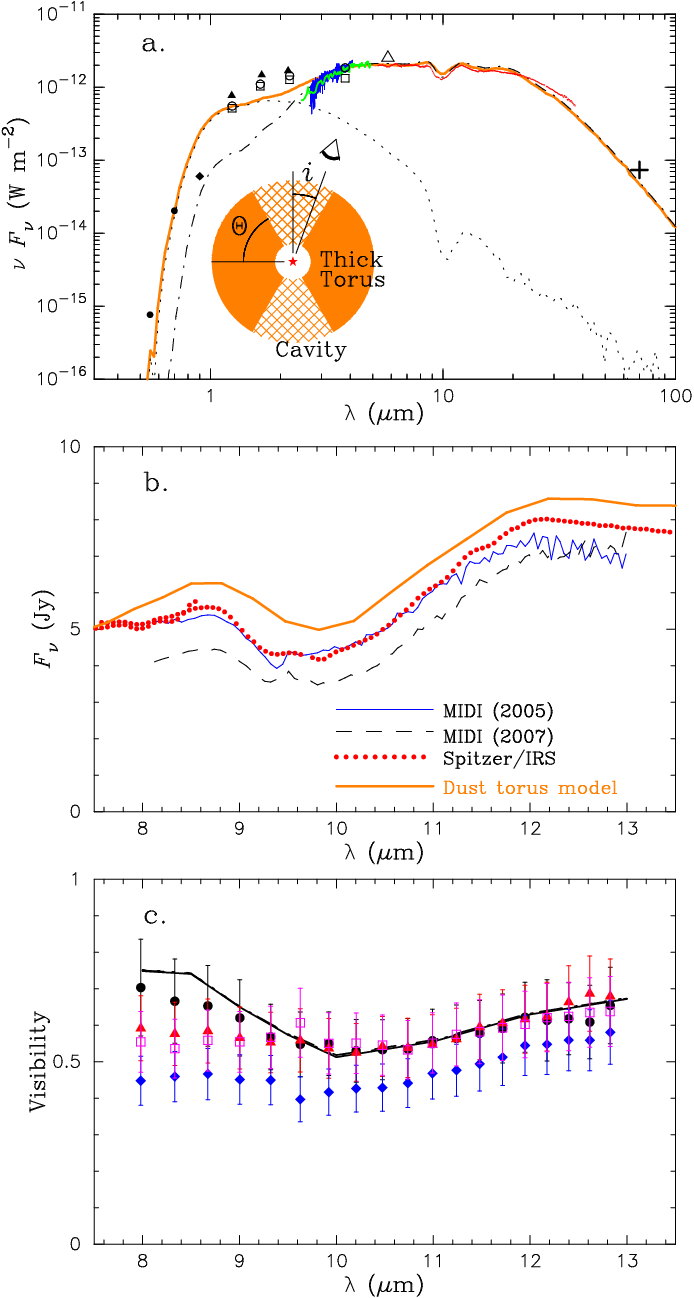}}}
\caption{Best-fit dust torus model for WOH G64.  
As shown in the inset of the panel {\bf a}, the torus is geometrically and 
optically thick and viewed relatively close to pole-on (see 
Table~\ref{table_res} for the parameters).  
{\bf a:} Comparison of the observed and model SEDs.  The SED of the torus 
model is plotted by the thick solid line.  The symbols as well as the blue, 
green, and red lines represent the observed SED (see Fig.~\ref{obs_sed}a for 
the references of the symbols).  
The dotted line represents the stellar flux contribution of the torus model.  
A spherical shell model is plotted with the dashed-dotted line.  
This spherical model has \TAUV\ = 20, and the other parameters are the same 
as the best-fit torus model.  
{\bf b:} Comparison of the mid-infrared spectra.  
{\bf c:} The observed $N$-band visibilities are plotted by the filled 
circles (data set \#2), filled triangles (\#4), filled diamonds (\#5), and 
open squares (\#6).  
The solid, dashed, and dotted lines (almost entirely overlapping with one 
another) represent the model visibilities calculated for different
orientations of the torus symmetry axis 
(solid line: torus axis perpendicular to the baseline vector, 
dashed line: 45\degr, dotted line: parallel), 
as illustrated in Fig.~\ref{best_image}. 
These model visibilities show no notable difference, because the model is 
viewed close to pole-on.  
}
\label{bestmodel}
\end{figure}

\begin{figure*}
\resizebox{\hsize}{!}{\includegraphics{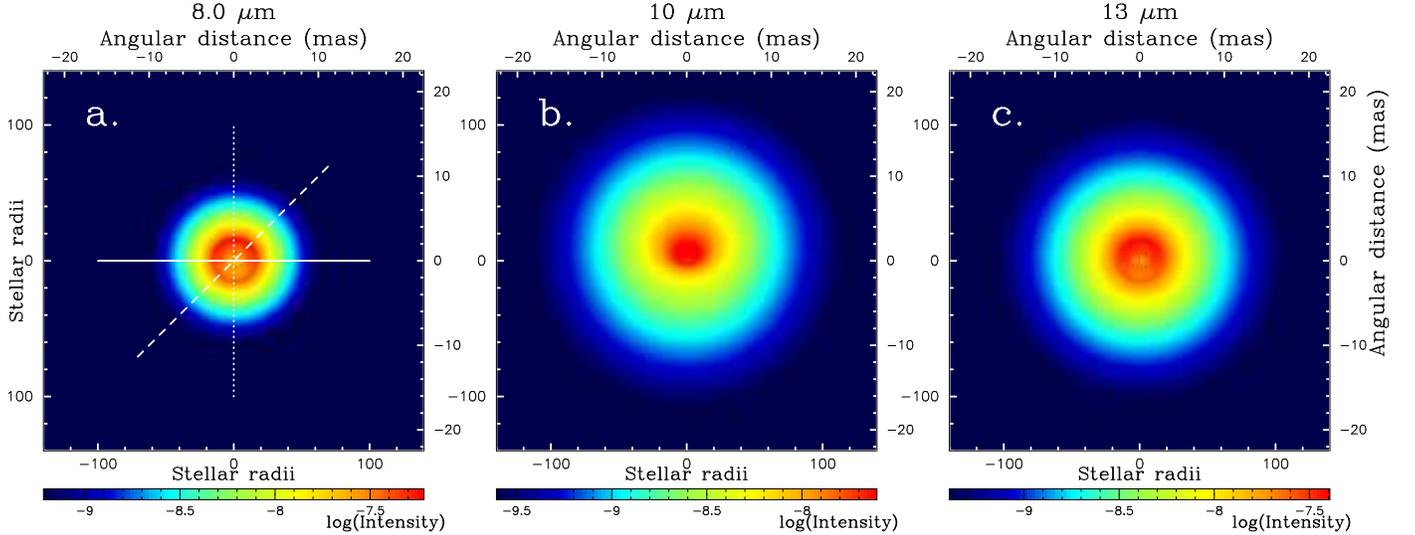}}
\caption{
  $N$-band images predicted by the best-fit torus model for WOH G64 shown in 
  Fig.~\ref{bestmodel}.  
  The color scale is normalized with the maximum intensity (excluding the
  star) and the minimum intensity set to 1\% of the maximum intensity.  
  The inner rim of the torus can be seen as the bright (red) circle on the 
  8~\micron\ image.  
  The orientations of the baseline vector used for the 
  calculation of the visibilities in Fig.~\ref{bestmodel}c are shown by 
  the solid, dashed, and dotted lines.  
}
\label{best_image}
\end{figure*}

The torus model considered here is similar to that adopted by Ohnaka 
et al. (\cite{ohnaka06}), which is basically a sphere with a biconical 
cavity.  We introduce two differences compared to this previous model: 
1) the density in the cavity is lower than in the torus but not exactly 
zero, and 2) the density drop from the torus to the cavity (along the 
latitudinal direction) is smoothed as described by 
\[
\rho(r, \theta) \propto \left( \frac{\RIN}{r} \right)^{p} \times \left(
\frac{1}{e^{( | \theta |  - \Theta)/\varepsilon} + 1} 
+ \frac{f}{e^{-(| \theta |  - \Theta)/\varepsilon} + 1} \right), 
\]
where $r$ and $\theta$ are the radial distance from the central star 
and the latitudinal angle measured from the equatorial plane, respectively, 
while \RIN\ and $\Theta$ are the inner boundary radius and the half-opening 
angle of the torus, respectively (see the inset of Fig.~\ref{bestmodel}a).  
$\varepsilon$ is a constant representing 
the smoothing of the density drop from the torus to the bipolar cavity.  
In the models discussed below, we adopt $\varepsilon = 0.03$ (for 
$\theta$ and $\Theta$ measured in radian).  
$f$ represents the ratio of the density in the cavity to that in the torus. 
The radial density distribution is assumed to be proportional to $r^{-2}$ 
(i.e., $p = 2$), 
which corresponds to a steady mass outflow with a constant velocity.  
The free parameters in our models are \RIN, $\Theta$, \TAUV\ (the radial 
optical depth at 0.55~\micron\ in the equatorial plane), and $f$.  
In addition to 
these parameters describing the torus, it is necessary to specify its 
orientation in the plane of the sky.  This is equivalent to 
specifying the angle between the torus symmetry axis projected onto 
the plane of the sky and the projected baseline vector of our MIDI 
observations (see also Fig.~\ref{best_image}), and this angle is used as an 
additional free parameter.  
The torus is assumed to consist of amorphous silicate grains, because 
the MIDI, Spitzer/IRS, and ISO spectra show no obvious dust features 
other than the self-absorption due to amorphous silicate.  
The grain size distribution is proportional to $a^{-q}$ between the minimum 
size (\AMIN) and maximum size (\AMAX), as considered by 
Mathis et al. (\cite{mathis77}). We tentatively assume \AMIN\ = 0.005~\micron, 
\AMAX\ = 0.1~\micron, and $q$ = 3.5, and examine the effects of these 
parameters later.  
The mean of the absorption and scattering coefficients weighted with the 
above size distribution is computed from the complex refractive index of 
the ``warm silicate'' of Ossenkopf et al. (\cite{ossenkopf92}) using the Mie 
code of Bohren \& Huffman (\cite{bohren83}).  For simplicity, isotropic 
scattering is considered for our modeling.  

The range of the free parameters varied in our modeling is as follows: 
\RIN\ (\RSTAR) = 5 ... 25 with $\Delta \RIN$ = 5, $\Theta$ = 20\degr\
... 90\degr\ with $\Delta \Theta$ = 10\degr\ ($\Theta$ = 90\degr\ corresponds 
to spherical shell models), \TAUV\ = 10 ... 50 with $\Delta 
\TAUV$ = 5,  and $f$ = 0.0 ... 0.5 with $\Delta f$ = 0.1.  
The outer boundary radius (\ROUT) is fixed to 250$\times$\RIN\ (see 
discussion below).  
SEDs are computed for different inclination angles (measured from pole-on 
as illustrated in Fig.~\ref{bestmodel}a) of 
$i$ = 0\degr ... 90\degr\ with $\Delta i$ = 10\degr\ for all models, and 
we first compare these model SEDs with the observed one.  
This comparison between the model and observed SEDs allows us to derive 
the bolometric luminosity of the central star, because the distance to 
the LMC is known (50~kpc).  
The two-dimensional visibilities are then computed from the Fourier 
transform of model $N$-band images. 
Note that the stellar radius derived from the central star's luminosity 
and effective temperature is used to scale the model visibilities.  

Figure~\ref{bestmodel} shows a comparison of the best-fit model 
with the observed SED and $N$-band visibilities, and the parameters of 
the model are given in Table~\ref{table_res}.  
As shown in the inset of Fig.~\ref{bestmodel}a, the best-fit model 
is characterized by a geometrically and optically thick torus viewed 
close to pole-on.  
The whole SED including the 10~\micron\ silicate feature is reasonably 
reproduced.  
The torus is optically thick in the equatorial direction (\TAUV\ = 30 
and $\tau_{\rm 10\mu m}$ = 7), which gives rise to the huge mid- and 
far-infrared excess as observed.  
However, since the object is seen through the cavity with lower 
density (\TAUV\ = 9 and $\tau_{\rm 10\mu m}$ = 2.1), the central 
star is not entirely obscured, and the flux in the near-infrared 
and in the optical as well as the observed silicate absorption is 
reasonably reproduced.  
The model predicts the near-infrared ($H$ and $K$ bands) flux 
to be somewhat lower than the observation.  This is because 
the spectrum of the central star is approximated with that of a 
blackbody.  The spectrum of photospheric models shows a flux excess 
at these wavelengths compared to the blackbody of the same effective 
temperature, because the opacity of negative hydrogen, 
which is the major continuous opacity source in the atmosphere of cool stars, 
has a minimum at $\sim$1.6~\micron\ (e.g., Tsuji \cite{tsuji78}; 
Plez et al. \cite{plez92}).  
The discrepancy in the optical flux between the 
model and the observation can be explained by molecular absorption 
features (mainly TiO) which are not included in the blackbody 
approximation.  
For comparison, we show a spherical shell model with the same parameters 
as the best-fit torus model, but with \TAUV\ = 20 (and a higher luminosity of 
$3.8 \times 10^5$~\LSOL\ than the $2.8 \times 10^5$~\LSOL\ of the torus model 
discussed below).    
This spherical shell model can reproduce the mid- and far-infrared excess 
as well as the 10~\micron\ silicate feature, but the predicted near-infrared 
and visible fluxes are too low to explain the observed data, because the 
optical depth along the line of sight is much higher.  

The model visibilities are also in agreement with the MIDI observations 
within the errors of the measurements.  
This pole-on model appears nearly centrosymmetric as seen in the 
$N$-band images shown in Fig.~\ref{best_image}, and this can explain 
the absence of a noticeable position-angle dependence in the observed 
visibilities.  
The orientation of the torus in the plane of the sky cannot be well 
constrained because the object is seen close to pole-on, but obviously, 
it is not a crucial parameter in such a case and 
does not affect the determination of the other parameters.  
We note that the inner boundary radius is weakly constrained, 10--20~\RSTAR, 
because the four visibilities currently available were taken at approximately 
the same baseline length and cannot put tight constraints on the radial 
density distribution.  This hampers us from discussing possible 
metallicity effects on dust formation in LMC RSGs at the moment.

We study the effects of the outer boundary radius by varying \ROUT\ 
in the best-fit model.  
While the adoption of larger outer boundary radii (\ROUT\ = 500~\RIN\ and 
1000~\RIN) does not result in a notable change in the SED below 70~\micron\ 
and $N$-band visibilities, 
the models with the outer boundary radii smaller than 250~\RIN\ 
predict the 70~\micron\ flux to be too low compared to the Spitzer/MIPS 
data.  
Therefore, we can only set a lower limit of 250$\times$\RIN\ on the outer 
boundary radius.  
We also check the effects of the grain size distribution by changing 
the minimum and maximum grain sizes as well as the power-law index 
of the size distribution for the best-fit model.  
We find the upper limit for the minimum grain size to be 0.05~\micron, 
while the lower limit cannot be well constrained.  
On the other hand, the upper and lower limits for the maximum grain size 
are found to be $\sim$0.15~\micron\ and $\sim$0.01~\micron, respectively.  
Models with power-law index values between 1.5 and 4.5 (3.5 is adopted 
in the above modeling) can reproduce the observational data, and thus, this 
parameter is not tightly constrained by the present modeling.

\begin{table}[h]
\begin{center}
\caption {
The parameters of the circumstellar dust torus and the central star of WOH G64 
derived from our radiative transfer modeling.  
}
\begin{tabular}{l l }\hline
Parameter  &  Value  \\ \hline
Torus half-opening angle ($\Theta$) &  60\degr$\pm$10\degr \\
Torus inner boundary radius  (\RIN) &  $15 \pm 5$~\RSTAR\ \\
(with 1~\RSTAR\ = 8~AU)           &  ($120 \pm 40$~AU) \\
Inner boundary dust temperature      &  $880^{+190}_{-110}$~K \\
Torus outer boundary radius (\ROUT) &  $\ge 250 \, \times$ \RIN \\
Radial optical depth at 0.55~\micron\ (\TAUV) & $30 \pm 5$ \\
Optical depth along the line of sight &   $9 \pm 2$ ($f = 0.3 \pm 0.1$) \\
Radial density distribution          & $r^{-2}$ (fixed) \\
Inclination angle ($i$)              & 20\degr\ (0\degr -- 30\degr\ with 
$i < 90\degr - \Theta$) \\

Grain size distribution              &  $\propto a^{-3.5}$ \\
Maximum grain size (\AMAX)           &  0.1~\micron\ (0.01--0.15~\micron) \\
Minimum grain size (\AMIN)           &  0.005~\micron\ ($<$0.05~\micron) \\

Total envelope mass                  &  3--9~\MSOL \\
                                     &  (gas-to-dust ratio = 200--500) \\

Luminosity                           &  $(2.8 \pm 0.3) \times 10^5$~\LSOL \\
Stellar radius                       &  1730~\RSOL \\ 
Effective temperature                &  3200~K (fixed) \\ \hline

\label{table_res}
\end{tabular}
\end{center}
\end{table}

\section{Discussion}
\label{sect_discuss}

Our two-dimensional modeling of the MIDI data and SED indicates that WOH G64 
has an axisymmetric, optically thick dust torus.  
The presence of such axisymmetric structures is suggested toward Galactic RSGs 
as well.  In particular, the well-studied dusty RSG, NML~Cyg, which can be 
regarded as a Galactic counterpart of WOH G64 
(van Loon et al. \cite{vanloon98}), is also suspected to have bipolar outflows 
based on radio interferometry observations (Richards et al. \cite{richards96}).  
Therefore, the occurrence of bipolar outflows and/or thick tori seems to 
be a common phenomenon among RSGs in the Galaxy as well as in the LMC, 
although the origin of the axisymmetric structures is still unclear.  
We note that the pole-on dusty torus model for WOH G64 is qualitatively 
consistent with the asymmetric OH maser profile with the blue peak much 
stronger than the red peak.  Marshall et al. (\cite{marshall04}) propose that 
the amplification of the stellar light is responsible for the blue-red-asymmetry, 
and this process can be particularly effective, when the star is seen through the 
less dense cavity as in the case of WOH G64.

The luminosity of the central star derived from our radiative transfer 
modeling, $2.8 \times 10^5$~\LSOL\ (\MBOL\ = $-8.8$), is by a factor 
of $\sim$2 lower than the (5--6)$\times 10^5$~\LSOL\ previously estimated by 
Elias et al. (\cite{elias86}) and van Loon et al. (\cite{vanloon05}) 
assuming a spherical shell.  Our modeling suggests 
that the object has an optically thick torus and is seen through the cavity 
with lower density.  
In such a case, more radiation escapes toward the cavity (i.e., 
toward the observer) than toward the optically thick torus, 
and the bolometric luminosity of the central star is overestimated, when it is 
derived from the observed flux assuming isotropic radiation.  
The location of WOH G64 based on the new, lower luminosity is plotted in 
Fig.~\ref{hr_diagram}.  
The newly derived luminosity corresponds to that of an RSG 
with an initial mass of $\sim$25~\MSOL, and the location of WOH G64 is in 
much better agreement with the theoretical evolutionary tracks, although 
WOH G64 is still somewhat too cool compared to the theoretical models.  
In fact, comparison with the Hayashi limit presented in 
Levesque et al. (\cite{levesque07}) 
reveals that 
WOH G64 is to the right of the Hayashi limit if \TEFF\ is 3200~K.  
Even if we adopt \TEFF\ = 3400~K corresponding to the earlier spectral type 
of M5 suggested by van Loon et al. (\cite{vanloon05}), it is close to, but 
still slightly beyond the Hayashi limit.  
The extreme effective temperature of WOH G64 suggests 
that it may be experiencing a very unstable phase accompanied by heavy 
mass loss, as proposed for 
some RSGs in the LMC and SMC located in the forbidden zone beyond 
the Hayashi limit (Levesque et al. \cite{levesque07}). 
However, it should also be kept in mind that the determination of effective 
temperature from the optical spectrum may be affected by the extinction due 
to the circumstellar dust envelope and/or the contribution of scattered light 
by dust grains, as suggested from the redshifted photospheric absorption lines 
with respect to the SiO and \HOH\ masers (van Loon et al. \cite{vanloon98}, 
\cite{vanloon01}).  
Also, the Hayashi limit is a complex function of stellar mass 
and metallicity as described in Cox \& Giuli (\cite{cox68}).  Therefore, it 
is still possible that WOH G64 is just on the Hayashi limit with an effective 
temperature somewhat higher than the above estimates.

WOH G64 indeed appears to have already shed a significant fraction of its 
initial mass.  
The dust mass in the best-fit torus model is $\sim 1.7 \times 10^{-2}$~\MSOL\ 
with a grain bulk density of 3~g~cm$^{-3}$, which translates into a total 
envelope mass of $\sim$3--9~\MSOL\ with gas-to-dust ratios of 200--500.  
These values already represent a noticeable fraction of the initial total 
mass of $\sim$25~\MSOL\ estimated for WOH G64.  
Moreover, as discussed in Sect.~\ref{sect_modeling}, the outer boundary 
radius of 250$\times$\RIN\ is a lower limit, and the envelope mass becomes 
even higher for larger outer boundary radii.  
Interestingly, although WOH G64 is cooler than the evolutionary tracks for 
a 25~\MSOL\ star shown in Fig.~\ref{hr_diagram}, the evolutionary 
calculation from Schaerer et al. (\cite{schaerer93}) as well as the new 
calculation performed in this work predicts that a 25~\MSOL\ star 
has lost 2 and 7.5~\MSOL\ at the RSG stage (an age of 7.8~Myr), respectively.  
Therefore, the derived envelope mass of WOH G64 is in rough agreement with 
current stellar evolution theory.

\section{Concluding remarks}

We have spatially resolved the circumstellar environment of the dusty 
red supergiant WOH G64 in the LMC, which is the first MIDI observation 
to resolve an individual stellar source in an extragalactic system.  
The observed $N$-band uniform-disk diameter is found to increase from 
18~mas at 8~\micron\ to 26~mas at 13~\micron.  
The visibilities obtained at approximately the same baseline length but 
at four position angles differing by $\sim$60\degr\ show no remarkable 
difference, indicating that the object appears nearly centrosymmetric.  
Our two-dimensional radiative transfer modeling shows that 
these visibilities and the observed SED can be explained by an optically and 
geometrically thick torus viewed close to pole-on.  
The luminosity of the central star derived by our modeling is by a factor 
of $\sim$2 lower than the previous estimates based on spherical shell models. 
The new, lower luminosity brings WOH G64 in much better agreement with 
the theoretical evolutionary tracks, which suggests an initial mass of 
$\sim$25~\MSOL\ for WOH G64.   
However, its location on the H-R diagram is very near or beyond 
the Hayashi limit.  
This implies that WOH G64 may be in an unstable stage at the moment, 
experiencing remarkable mass loss.  The total envelope mass estimated from 
our model is of the order of 3--9~\MSOL, representing a significant fraction 
of its initial mass.  This envelope mass is comparable to what is predicted 
by current stellar evolutionary calculations.

We also identified the \HOH\ absorption features at 2.7~\micron\ and 
6~\micron\ in the spectra obtained with ISO and Spitzer, respectively.  
The 2.7~\micron\ feature is likely to originate in the photosphere and/or 
MOLsphere.  
On the other hand, the 6~\micron\ feature can be interpreted as the 
absorption due to cold circumstellar \HOH\ gas seen with the hot dust thermal 
emission as background.

Obviously, follow-up MIDI observations are necessary for putting 
more constraints on the properties of the torus.  
In particular, MIDI observations with shorter and longer baselines are 
indispensable for obtaining tighter constraints on the torus inner boundary 
radius, which is crucial for better understanding possible metallicity effects 
on the dust formation and mass loss in RSGs.  
MIDI measurements over even a wider position angle range are also important 
for probing the possible presence of more complex structures.  
The \HOH\ gas in the photosphere and in the MOLsphere 
can be studied in more detail using the near-infrared VLTI instrument 
AMBER (Petrov et al. \cite{petrov07}).  
For example, AMBER observations of the \HOH\ bands in the $K$ band would be 
useful for constraining the geometrical extension of the \HOH\ layers 
close to the star.

\begin{acknowledgement}

We thank the ESO VLTI team on Paranal and in Garching and the MIDI 
team for carrying out the observations and making the data reduction 
software publicly available.  
This paper utilizes public domain data originally obtained by the
MACHO Project, whose work was performed under the joint auspices of
the U.S. Department of Energy, National Nuclear Security
Administration by the University of California, Lawrence Livermore
National Laboratory under contract No. W-7405-Eng-48, the National
Science Foundation through the Center for Particle Astrophysics of the
University of California under cooperative agreement AST-8809616, and
the Mount Stromlo and Siding Spring Observatory, part of the
Australian National University.

\end{acknowledgement}

\end{document}